\documentstyle [12pt] {article}
\begin{document}
\setcounter{page}{1}
\def\theequation{\arabic{section}.\arabic{equation}}
\def\theequation{\thesection.\arabic{equation}}
\setcounter{section}{0}

\title{On the Wilson loop in the dual representation within the dual
Higgs model with dual Dirac strings}

\author{V. A. Ivanova\,\thanks{Permanent Address: State Technical
University, Department of Nuclear Physics, 195251 St. Petersburg,
Russian Federation}~~and~ N. I. Troitskaya~${^*}$}

\date{\today}

\maketitle
\vspace{-0.5in}
\begin{center}
{\it Atominstitut der \"Osterreichischen Universit\"aten,
Arbeitsbereich Kernphysik und Nukleare Astrophysik, Technische
Universit\"at Wien, \\ Wiedner Hauptstr. 8-10, A-1040 Wien,
\"Osterreich }
\end{center}

\begin{center}
\begin{abstract}
The vacuum expectation value of the Wilson loop in the dual
representation is calculated in the dual Higgs model with dual Dirac
strings.  It is shown that the averaged value of the Wilson loop in
the dual representation obeys the area--law falloff. Quantum
fluctuations of the dual--vector and the Higgs field around Abrikosov
flux lines induced by dual Dirac strings in a dual superconducting
vacuum and string shape fluctuations are taken into account.
\end{abstract}
\end{center}

\newpage

\section{Introduction}
\setcounter{equation}{0}

\hspace{0.2in} A dual $U(1)$ gauge Higgs model with dual Dirac strings
suggested in Ref.[1] has been then applied to the calculation of the
effective string energy caused by quantum field and string shape
fluctuations around an Abrikosov flux line stretched between a quark
and an antiquark [2,3]. In the symmetry broken phase due to a
superconducting vacuum a dual Dirac string induces a dual--vector
field $C_{\mu}(x; X)$ with a shape of an Abrikosov flux line. It has
been shown that quantum field fluctuations give the contribution to
the string tension [2], while the string shape fluctuations induce 
Coulomb--like contributions [4] with an universal coupling constant
$\alpha_{\rm string }=\pi/12$ and $\alpha_{\rm string }=\pi/3$ for
open and closed strings, respectively [3].

In this paper we turn to the investigation of the Wilson loop [5] in
the dual representation. In the dual Higgs model with $U(1)$ gauge
symmetry the Wilson loop is determined by
\begin{eqnarray}\label{label1.1}
W({\cal C}) = \exp\Big\{ i\,g\, \oint_{\cal C}d
\,x^{\mu}\,C_{\mu}(x)\Big\},
\end{eqnarray}
where $g$ is a gauge coupling constant and ${\cal C}$ is a closed
contour. 

In the symmetric phase the Lagrangian of the dual Higgs model with
dual Dirac strings is defined by [1]
\begin{eqnarray}\label{label1.2}
{\cal L}(x) &=& \frac{1}{4}F_{\mu\nu}(x)F^{\mu\nu}(x) +
(\partial_{\mu}+igC_{\mu}(x))\Phi^{\ast}(x)(\partial^{\mu} -ig
C^{\mu}(x))\,\Phi(x)\nonumber\\ &&-\kappa^2
(v^2-\Phi^{\ast}(x)\Phi(x))^2 + {\cal L}_{\rm free~quark}(x),
\end{eqnarray}
where ${\cal L}_{\rm free~quark}(x)$ is the kinetic term of 
classical quarks and antiquarks [1].

Then $\Phi(x)$ is a complex Higgs field with a vacuum expectation
value $v$, $\langle \Phi \rangle = v$, and $\kappa$ is the coupling
constant. The field strength $F^{\mu\nu}(x)$ is defined by:
$F^{\mu\nu}(x) = {\cal E}^{\mu\nu}(x) - {^* C^{\mu\nu} (x)}$, where
$C^{\mu\nu}(x) = \partial^{\mu} C^{\nu}(x) - \partial^{\nu}
C^{\mu}(x)$, and ${^*C^{\mu\nu} (x)} =
\frac{1}{2}\,\varepsilon^{\mu\nu\alpha\beta} C_{\alpha\beta} (x)$ is a
dual version of $C^{\mu\nu} (x)$ [1]. The electric field strength
${\cal E}^{\mu\nu}(x)$ is induced by a dual Dirac string and
determined as follows [1--3]
\begin{eqnarray}\label{label1.3}
{\cal E}^{\mu\nu} (x) &=& Q \int\!\!\!\int_{S(L)} d\sigma^{\mu\nu}(X)
\delta^{(4)}(x - X) =\nonumber\\ &=&Q \int\!\!\!\int_{S(L)}
d\tau\,d\sigma\,\sigma^{\mu\nu}(X) \delta^{(4)}(x - X),
\end{eqnarray}
where $S(L)$ is a 2-dimensional surface swept on the world--sheet by a
shape $L$ of a dual Dirac string and $Q$ is the electric charge of a
quark. The surface is parameterized by internal variables ($-\infty <
\tau < + \infty$ and $0 \le \sigma \le \pi$) [1--3]:
\begin{eqnarray}\label{label1.4}
d\sigma^{\mu\nu}(X) =\sigma^{\mu\nu}(X) d\tau d\sigma =
\Bigg(\frac{\partial X^{\mu}}{\partial \tau}\,\frac{\partial
X^{\nu}}{\partial\,\sigma} - \frac{\partial X^{\nu}}{\partial
\tau}\,\frac{\partial X^{\mu}}{\partial \sigma}\Bigg) d\tau d\sigma,
\end{eqnarray}
such as $X^{\mu}(\tau,\sigma)|_{\sigma=0} = X^{\mu}_{\bar{q}}(\tau)$,
$X^{\mu}(\tau,\sigma)|_{\sigma=\pi} = X^{\mu}_q(\tau)$ represent the
world lines of an antiquark and a quark, respectively. As has been
shown in [1] the inclusion of a dual Dirac string in terms of ${\cal
E}^{\mu\nu}(x)$ defined by (\ref{label1.3}) saturates the electric
Gauss law. Below in order to underscore that ${\cal E}^{\mu\nu}(x)$ is
a functional of $X$, a point upon the surface $S(L)$, we introduce the
notations ${\cal E}^{\mu\nu}(x) \to {\cal E}^{\mu\nu}(x; X)$ and
${\cal L}_{\rm free~quark}(x)\to {\cal L}_{\rm free~quark}(x; X)$.

In the symmetry--broken phase it is convenient to use a polar
representation of the Higgs field $\Phi$, i.e.  $\Phi(x) =
\rho(x)\,e^{\textstyle i\,\vartheta(x)}$. In the polar representation
of the Higgs field the Lagrangian (\ref{label1.2}) reads
\begin{eqnarray}\label{label1.5}
{\cal L}(x) &=& \frac{1}{4}F_{\mu\nu}(x)F^{\mu\nu}(x) +
(\partial_{\mu} + ig{\tilde
C}_{\mu}(x))\rho(x)(\partial^{\mu}-ig{\tilde C}^{\mu}(x))\rho(x) -
\nonumber\\ &&- \kappa^2 (v^2 - \rho^2(x))^2 + {\cal L}_{\rm
free~quark}(x; X),
\end{eqnarray}
where we have denoted ${\tilde C}_{\mu}(x) = C_{\mu}(x) -
\partial_{\mu}\vartheta(x)/g$. Below we omit a tilde above the
$C_{\mu}$--field.

The singular part of the phase field $\theta(x)$ and the electric
tensor field ${\cal E}^{\mu\nu}(x; X)$ are related by [2]: ${^*\!{\cal
E}^{\mu\nu}(x; X)} = (1/g) (\partial^{\mu} \partial^{\nu} -
\partial^{\nu} \partial^{\mu}) \theta(x)$ (see also [6]).

The transition to the symmetry--broken phase can be performed by the
shift of the $\rho$--field: $\rho(x) = v + \sigma(x)/\sqrt{2}$. This
reduces the Lagrangian (\ref{label1.5}) to the form
\begin{eqnarray}\label{label1.6}
&&{\cal L}(x) = \quad\frac{1}{4} F_{\mu\nu}(x)\,F^{\mu\nu}(x) +
\frac{1}{2} M^2_C C_{\mu}(x) C^{\mu}(x) + \nonumber\\ &&+g M_C
{\sigma}(x) \Bigg[1 +
\frac{\kappa}{\sqrt{2}}\,\frac{\sigma(x)}{M_{\sigma}}\Bigg] C_{\mu}(x)
C^{\mu}(x) +
\frac{1}{2}\partial_{\mu}\sigma(x)\,\partial^{\mu}\sigma(x) - \\ &&-
\frac{1}{2}\,M^2_{\sigma}\,{\,\sigma}^2(x)\,\Bigg[1 +
\frac{\kappa}{\sqrt{2}}\,\frac{\sigma(x)}{M_{\sigma}}\Bigg]^2 + {\cal
L}_{\rm free~quark}(x),\nonumber
\end{eqnarray}
where $M_C = \sqrt{2}gv$ and $M_{\sigma} = 2 \kappa v $ are the masses
of $C_{\mu}$ and $\sigma$ fields, respectively.

\section{The averaged value of the Wilson loop in the dual 
representation}
\setcounter{equation}{0}

\hspace{0.2in} In the symmetry--broken phase the dual--vector field
$C_{\mu}$ fluctuates around an Abrikosov flux line induced by a dual
Dirac string in a superconducting vacuum [1--3]. By varying the
Lagrangian (\ref{label1.6}) with respect to the field $C_{\nu}(x)$ and
using the constraint $\partial_{\nu}C^{\nu}(x) = 0$ we get the
equation of motion [1--3]
\begin{eqnarray}\label{label2.1}
(\Box + M^2_C)\, C^{\nu}(x) = - \partial_{\mu}{^*{\cal E}}^{\mu\nu}(x;
X) - 2 g M_C \sigma (x)\,\Bigg[1 +
\frac{\kappa}{\sqrt{2}}\frac{\sigma(x)}{M_{\sigma}}\Bigg] C^{\nu}(x).
\end{eqnarray}
In the tree approximation the vacuum expectation value $\langle
C^{\nu}(x; X) \rangle$ obeys the equation [1--3]
\begin{eqnarray}\label{label2.2}
(\Box + M^2_C)\,\langle C^{\nu}(x; X) \rangle = -
\partial_{\mu}{^*{\cal E}}^{\mu\nu}(x; X),
\end{eqnarray}
where we have taken into account that in the tree approximation
$\langle \sigma(x) \rangle=0$. The solution of Eq.(\ref{label2.2}) has
the shape of a dual Abrikosov flux line and reads [1--3]
\begin{eqnarray}\label{label2.3}
\langle C^{\nu}(x; X) \rangle = - \int d^4x^{\prime}\,\Delta\,(x -
x^{\prime},M_C)\,\partial_{\mu}{^*{\cal E}}^{\mu\nu}(x^{\prime}; X),
\end{eqnarray}
where $\Delta(x - y, M_C)$ is the Green function
\begin{eqnarray}\label{label2.4}
\Delta (x - y, M_C) = \int
\frac{d^4k}{(2\pi)^4}\frac{1}{M^2_C - k^2 - i\,0}\,e^{\displaystyle
-ik\cdot (x - y)}.
\end{eqnarray}
The Abrikosov flux line provides a linearly rising interquark
potential realizing confinement of electric quark charges [1--3].

The calculation of the averaged value of the Wilson loop $W({\cal C})$
we perform integrating over the dual--vector field $c_{\mu}(x)$,
$C_{\mu}(x) = \langle C_{\mu}(x; X) \rangle + c_{\mu}(x)$, and the
Higgs field $\sigma(x)$ fluctuating around the Abrikosov flux line and
obeying the constraints $\langle c_{\mu}(x) \rangle = 0$ and $\langle
\sigma(x) \rangle = 0$ [2]. The fields $c_{\mu}(x)$ and $\sigma(x)$
are described by the Lagrangian
\begin{eqnarray}\label{label2.5}
{\cal L}(x) &=&{\cal L}_{\rm string}(x; X) \nonumber\\
&+&\frac{1}{2}\,c_{\mu}(x)\Bigg\{\Bigg(\Box + M^2_C + g\,M_C
{\sigma}(x)\,\Bigg[1 +
\frac{\kappa}{\sqrt{2}}\,\frac{\sigma(x)}{M_{\sigma}}\Bigg]\Bigg)\,
g^{\mu\nu} - \partial^{\mu} \partial^{\nu}\Bigg\}\,c_{\nu}(x)
\nonumber\\ &+&2\,g\,M_C {\sigma}(x)\,\Bigg[1 +
\frac{\kappa}{\sqrt{2}}\,\frac{\sigma(x)}{M_{\sigma}}\Bigg]\,\langle
C_{\mu}(x; X) \rangle\,c^{\mu}(x) \nonumber\\ &+&g\,M_C
{\sigma}(x)\,\Bigg[1 +
\frac{\kappa}{\sqrt{2}}\,\frac{\sigma(x)}{M_{\sigma}}\Bigg]\,\langle
C_{\mu}(x; X) \rangle\,\langle C^{\mu}(x; X) \rangle \nonumber\\ &+&
\frac{1}{2}\partial_{\mu}\,\sigma(x)\,\partial^{\mu}\,\sigma(x) -
\frac{1}{2}\,M^2_{\sigma}\,{\sigma}^2(x)\,\Bigg[1\,+
\frac{\kappa}{\sqrt{2}}\,\frac{\sigma(x)}{M_{\sigma}}\Bigg]^2
+ {\cal L}_{\rm free~quark}(x, X),
\end{eqnarray}
which can be obtained from (\ref{label1.6}).  In (\ref{label2.5}) we
have used Eq.(\ref{label2.2}). The Lagrangian ${\cal L}_{\rm
string}(x; X)$ depends explicitly on the shape of a dual Dirac string
and is defined by [1--3]
\begin{eqnarray}\label{label2.6}
\int d^4x{\cal L}_{\rm string}(x; X) = \frac{1}{4}\,M^2_C\int \int
d^4x\, d^4y \,{\cal E}_{\mu\alpha}(x; X)\,\Delta^{\alpha}_{\nu}(x - y,
M_C)\,{\cal E}^{\mu\nu}(y; X),
\end{eqnarray}
where $\Delta^{\alpha}_{\nu}(x - y, M_C) = (g^{\alpha}_{\nu} +
2\,\partial^{\alpha}\,\partial_{\nu}/M^2_C) \,\Delta (x - y; M_C)$.

Using Stokes' theorem and relations above the Wilson loop in the dual
representation Eq.(\ref{label1.1}) can be recast into the form
\begin{eqnarray}\label{label2.7}
W({\cal C}) &=& \exp\Big\{i\,\frac{1}{2}\,\int d^4x\,{\cal
E}_{\mu\nu}(x; X)\,{^*{\cal G}^{\mu\nu}(x)}\Big\}\,
\exp\Big\{\frac{1}{2}\,i\,\int d^4x\,\langle C_{\mu\nu}(x; X)
\rangle\,{\cal G}^{\mu\nu}(x)\Big\}\nonumber\\ &\times&\exp\Big\{
-i\,\int d^4x\,c_{\nu}(x)\,\partial_{\beta}\,{\cal
G}^{\beta\alpha}(x)\Big\},
\end{eqnarray}
where $\langle C_{\mu\nu}(x; X) \rangle = \partial_{\mu}\langle
C_{\nu}(x; X) \rangle - \partial_{\nu}\langle C_{\mu}(x; X) \rangle$
and ${\cal G}^{\mu\nu}(x)$ is determined by
\begin{eqnarray}\label{label2.8}
\hspace{-0.5in}{\cal G}^{\mu\nu} (x) &=& g \int\!\!\!\int_{S({\cal
C})} d\sigma^{\mu\nu}(Y) \delta^{(4)}(x - Y) = g
\int\!\!\!\int_{S({\cal C})}
d\tau^{\prime}\,d\sigma^{\prime}\,\sigma^{\mu\nu}(Y) \delta^{(4)}(x -
Y).
\end{eqnarray}
A 2-dimensional surface $S({\cal C})$ supports on the closed contour
${\cal C}$, $\partial S({\cal C}) = {\cal C}$, and $Y^{\alpha} \equiv
Y^{\alpha}(\tau^{\prime}, \sigma^{\prime}\,)$ is a point of this
surface. One can show that the contribution of the first exponential
is equal to unity, since the exponent
\begin{eqnarray}\label{label2.9}
\frac{1}{2}\,\int d^4x\,{\cal E}_{\mu\nu}(x; X)\,{^*{\cal
G}^{\mu\nu}(x)},
\end{eqnarray}
when the Dirac relation $Q g = 2\pi$ is used, amounts to either $2\pi$
or zero.

Averaging the Wilson loop $W({\cal C})$ over fluctuations of the
quantum fields $c_{\mu}(x)$ and $\sigma(x)$ we obtain the quantity
depending only on string and quark--antiquark degrees of freedom
$\langle W({\cal C}) \rangle = W[X]$ defined by
\begin{eqnarray}\label{label2.10}
W[X]=\int {\cal D}c_{\mu}\,{\cal D}\sigma\,W({\cal
C})\,\exp\Big\{ i\int d^4x\,{\cal L}(x; X)\Big\}.
\end{eqnarray}
The Lagrangian ${\cal L}(x; X)$ is given by Eq.(\ref{label2.2}). The
integral over the dual--vector field $c_{\mu}(x)$ is Gaussian and can
be calculated explicitly. The result reads
\begin{eqnarray}\label{label2.11}
\hspace{-0.3in}W[X] &=& \exp\Big\{i\int d^4x\Big[{\cal L}_{\rm
string}(x; X) + {\cal L}_{\rm free~quark} + \frac{1}{2} i \langle
C_{\mu\nu}(x; X) \rangle {\cal G}^{\mu\nu}(x)\Big]\Big\}\nonumber\\
&\times&\int {\cal D}\sigma\,\exp\Big\{i\int d^4x\,{\cal L}_{\rm
eff}(x; X)\Big\}.
\end{eqnarray}
The effective Lagrangian ${\cal L}_{\rm eff}(x; X)$ amounts to 
\begin{eqnarray}\label{label2.12}
&&\int d^4x\,{\cal L}_{\rm eff}(x; X) =-i\frac{1}{2}\int d^4x{\rm
tr}{\ell n} D^{-1}(x,y|\sigma)|_{y=x}\nonumber\\ &&+ \int
d^4x\Bigg\{\frac{1}{2}\partial_{\mu}\,\sigma(x)\,\partial^{\mu}\,
\sigma(x) -\frac{1}{2}\,M^2_{\sigma}\,{\sigma}^2(x)\,\Bigg[1\,+
\frac{\kappa}{\sqrt{2}}
\,\frac{\sigma(x)}{M_{\sigma}}\Bigg]^2\Bigg\}\nonumber\\ &&-
\frac{1}{2}\,\int\!\!\!\int d^4x\,d^4y\,{\cal J}^{\mu}(x; X)\,
D_{\mu\nu}(x, y|\sigma)\, {\cal J}^{\nu}(y; X) \nonumber\\ &&+ \int
d^4x\,g\,M_C\,\sigma(x)\,\Bigg[1 +
\frac{\kappa}{\sqrt{2}}\,\frac{\sigma(x)}{M_{\sigma}}\Bigg]\,\langle
C_{\mu}(x; X) \rangle\,\langle C^{\mu}(x; X) \rangle,
\end{eqnarray}
where we have denoted
\begin{eqnarray}\label{label2.13}
{\cal J}^{\alpha}(x; X) = \partial_{\beta}\,{\cal G}^{\beta\alpha}(x)
- 2\,g\,M_C {\sigma}(x)\,\Bigg[1 +
\frac{\kappa}{\sqrt{2}}\,\frac{\sigma(x)}{M_{\sigma}}\Bigg]\,\langle
C^{\alpha}(x; X) \rangle,
\end{eqnarray}
and $D_{\mu\nu}(x, y|\sigma)$ is the Green function of the
$c_{\mu}$--field in the external scalar field induced by the
self--interactions of the Higgs field $\sigma$
\begin{eqnarray}\label{label2.14}
i D_{\mu\nu}(x, y|\sigma) = \Bigg\langle 0\Bigg|{\rm
T}\Bigg(c_{\mu}(x) \, c_{\nu}(y)\, \exp{i\int d^4z\,{\cal L}_{\rm
int}(z)}\Bigg)\Bigg|0\Bigg \rangle,
\end{eqnarray}
where ${\rm T}$ is a time--ordering operator and
\begin{eqnarray}\label{label2.15}
{\cal L}_{\rm int}(z) = g\,M_C {\sigma}(z)\,\Bigg[1 +
\frac{\kappa}{\sqrt{2}}\,\frac{\sigma(z)}{M_{\sigma}}\Bigg]\,
c_{\mu}(z)\,c^{\mu}(z).
\end{eqnarray}
Making all convolutions of the $c_{\mu}$--fields we obtain
$D_{\mu\nu}(x, y|\sigma)$ as a functional of the
$\sigma$--field. Then, ${\rm tr}{\ell n}D^{-1}(x, y|\sigma)|_{y=x}$ in
Eq.(\ref{label2.12}) means the trace over Lorentz indices of the
inverse Green function of the $c_{\mu}(x)$--field $D^{-1}_{\mu\nu}(x,
y|\sigma)$ satisfying the equation
\begin{eqnarray}\label{label2.16}
&&D^{-1}_{\mu\nu}(x, y|\sigma) = \nonumber\\ &&=\Big\{ \Big(\Box_x +
M^2_C + g\,M_C {\sigma}(x)\,\Big[1 +
\frac{\kappa}{\sqrt{2}}\,\frac{\sigma(x)}{M_{\sigma}}\Big]\Big)\,
g_{\mu\nu} -\partial_{\mu}\partial_{\nu}\Big\}\,\delta^{(4)}(x - y).
\end{eqnarray}
The inverse Green function $D^{-1}_{\mu\nu}(x, y|\sigma)$ is connected
with the Green function $D_{\mu\nu}(x, y|\sigma)$ via the relation: $
\int d^4z\,D^{-1}_{\mu\lambda}(x, z|\sigma)\,{D^{\lambda}}_{\nu}(z,
y|\sigma) = g_{\mu\nu}\,\delta^{(4)}(x - y)$.

Integration over the $\sigma$--field is a very complicated problem and
cannot be carried out explicitly in a general form. In order to
integrate out the $\sigma$--field we suggest to apply the
approximation used in Refs. [1--3]: $M_{\sigma}\gg M_C$. This
corresponds to a strong coupling limit of the Higgs fields $\kappa \gg
g$. In this limit the scales of fluctuations of the $\sigma$--field
are small compared with $M_{\sigma}$ and only one--loop contributions
become dominant [1--3]. The result of the integration over the
$\sigma$--field reads
\begin{eqnarray}\label{label2.17}
\int {\cal D}\sigma\,\exp\Big\{ i\int d^4x\,{\cal
L}_{\rm eff}(x; X)\Big\} = \exp\Big\{ i\int
d^4x\,\delta\,{\cal L}_{\rm one-loop}(x; X)\Big\},
\end{eqnarray}
where we have denoted
\begin{eqnarray}\label{label2.18}
\hspace{-0.5in}&&\int d^4x\,\delta\,{\cal L}_{\rm one-loop}(x; X) = -
\frac{1}{2}\,g^2 i\,\Delta\,(0; M_{\sigma}) \int d^4x \langle
C_{\mu}(x; X) \rangle\langle C^{\mu}(x; X)
\rangle\nonumber\\\hspace{-0.5in}&& +2 i g^2 M^2_C \int\!\!\!\int d^4x
d^4y \langle C^{\mu}(x; X) \rangle \langle C^{\nu}(y; X) \rangle
D_{\mu\nu}(x - y; M_C)\Delta(y - x; M_{\sigma}).
\end{eqnarray}
Here $\Delta(x; M_{\sigma})$ is the Green function of the free
$\sigma$--field given by Eq.(\ref{label2.4}) by changing $M_C\to
M_{\sigma}$, then $D_{\mu\nu}(x; M_C) = (g_{\mu\nu}
+\partial_{\mu}\partial_{\nu}/M^2_C) \Delta(x; M_C)$ is the Green
function of the free $c_{\mu}$--field. We have also used the relations
$M_C=\sqrt{2} g v$ and $M_{\sigma} = 2 \kappa v$. Since $\langle
C^{\mu}(x; X) \rangle$ is proportional to the electric charge of a
quark $Q$, so due to the Dirac relation $Q\,g\,=\,2\pi$ [1--3] the
effective Lagrangian $\delta\,{\cal L}_{\rm one-loop}(x; X)$ does not
depend on $Q$. In turn the dependence of $\delta\,{\cal L}_{\rm
one-loop}(x; X)$ on the coupling constants $g$ or $\kappa$ enters
implicitly via the masses $M_C$ and $M_{\sigma}$. Using
Eq.(\ref{label2.18}) we obtain the averaged value of the Wilson loop
\begin{eqnarray}\label{label2.19}
\hspace{-0.3in}W[X] = \exp\Big\{i\int d^4x\Big[{\cal L}_{\rm
eff~string}(x; X) + {\cal L}_{\rm free~quark}(x; X) + \frac{1}{2}
\langle C_{\mu\nu}(x; X) \rangle {\cal G}^{\mu\nu}(x)\Big]\Big\},
\end{eqnarray}
where ${\cal L}_{\rm eff~string}(x; X) = {\cal L}_{\rm string}(x; X) +
\delta\,{\cal L}_{\rm one-loop}(x; X)$. The averaged value $W[X]$
given by Eq.(\ref{label2.19}) is a functional of a shape of a dual
Dirac string and quark--antiquark degrees of freedom.

\section{The averaged value of the Wilson loop for static 
quarks and strings} 
\setcounter{equation}{0}

\hspace{0.2in} First, we suggest to consider the calculation of the
r.h.s. of Eq.(\ref{label2.19}) for static quarks and strings. In this
case ${\cal L}_{\rm free~quark}(x; X) = 0$ and the r.h.s. of
Eq.(\ref{label2.19}) becomes equal to
\begin{eqnarray}\label{label3.1}
W[X] = \exp\Big\{i\int d^4x\Big[{\cal L}_{\rm
eff~string}(x; X) + \frac{1}{2} \langle C_{\mu\nu}(x; X) \rangle {\cal
G}^{\mu\nu}(x)\Big]\Big\}.
\end{eqnarray}
Following [1--3] we consider a static straight string of a length $L$
stretched along the $z$--axis between a quark and an antiquark placed
at the points $\vec{X}_Q = (0,0,L/2)$ and $\vec{X}_{ - Q} = (0,0,-
L/2)$, respectively. For such a static string the electric field
strength ${\cal E}_{\mu\nu}(x; X)$ does not depend on time and is
given by [1--3]
\begin{eqnarray}\label{label3.2}
\vec{\cal E}(\vec{x}, \vec{X}\,) = \vec{e}_z\,Q\,\delta (x)\,\delta
(y)\,\Big[\theta (z -\frac{1}{2}\,L) - \theta (z
+\frac{1}{2}\,L)\Big],
\end{eqnarray}
where the unit vector $\vec{e}_z$ is directed along the $z$--axis and
$\theta (z)$ is the Heaviside function. This field strength produces
the dual--vector potential
\begin{eqnarray}\label{label3.3}
\langle \vec{C}(\vec{x}, \vec{X}\,) \rangle = - \,i\,Q\,\int
\frac{d^3k}{4\,\pi^3}\,\frac{\vec{k} \times
\vec{e}_z}{k_z}\,\frac{1}{M^2_C + \vec{k}^{\,2}}\,\sin\Bigg(\frac{k_z
L}{2}\Bigg)\, {\displaystyle e^{\displaystyle
i\,\vec{k}\cdot\vec{x}}}.
\end{eqnarray}
In the Euclidean space--time the first term in the r.h.s. of
Eq.(\ref{label3.1}) can be represented in the form
\begin{eqnarray}\label{label3.4}
i\int d^4x\,{\cal L}_{\rm eff~string}(x; X) = - V(L)\,T,
\end{eqnarray}
where $T$ is a time and $V(L)$ is the interquark potential induced by
a dual Dirac string. Taking the limit $L\to \infty$, corresponding to
an infinitely long string, and keeping only the terms linear in $L$ we
obtain [2]
\begin{eqnarray}\label{label3.5}
V(L) &=&\frac{Q^2 M^2_C}{8 \pi}\,\Bigg(1 +
\frac{g^2}{16\pi^2}\frac{M^2_{\sigma}}{M^2_C}\Bigg)\,{\ell n}\Bigg(1 +
\frac{M^2_{\sigma}}{M^2_C}\Bigg)\,L = \nonumber\\ &=&
\pi\,v^2\,\Bigg(1 + \frac{\kappa^2}{8 \pi^2}\Bigg)\,{\ell n}\Bigg(1 +
\frac{\kappa^2 Q^2}{2 \pi^2}\Bigg)\,L = \sigma\,L,
\end{eqnarray}
where $\sigma$, the string tension, is determined
\begin{eqnarray}\label{label3.6}
\sigma =\pi\,v^2\,\Bigg(1 + \frac{\kappa^2}{8 \pi^2}\Bigg)\,{\ell
n}\Bigg(1 + \frac{\kappa^2 Q^2}{2\,\pi^2}\Bigg).
\end{eqnarray}
We have used here the relations $M_C = \sqrt{2} g v$, $M_{\sigma} = 2
\kappa v$ and $Q\,g = 2\pi$.

The second term ${\displaystyle \int d^4x\,\langle C_{\mu\nu}(x; X)
\rangle\, {\cal G}^{\mu\nu}(x)}$ in the exponent of (\ref{label3.3})
depends on the surface $S({\cal C})$ bounded by the contour ${\cal
C}$. One can show that for the shape of the string defined by
Eq.(\ref{label3.1}) this term vanishes for an infinitely long contour
${\cal C}$ embracing an infinitely large surface $S({\cal C})$. Thus,
the Wilson loop in the dual representation, calculated for static
quarks, antiquarks and a static straight string stretched between
them, acquires the form
\begin{eqnarray}\label{label3.7}
W[X] = {\displaystyle e^{\displaystyle - \sigma\,L\,T}},
\end{eqnarray}
where $L\,T$ is an area of a rectangular surface swept by a dual Dirac
string of a length $L$. Such a behaviour of the averaged value of the
Wilson loop given by Eq.(\ref{label3.7}) corresponds to the area--law
falloff [5] testifying confinement of electric quark charges [5,6].

\section{The averaged value of the Wilson loop for static quarks 
and fluctuating strings} 
\setcounter{equation}{0}

\hspace{0.2in} A dynamics of a dual Dirac string we take into account
considering string shape fluctuations for the string with static
quarks and antiquarks attached to the ends. Following [3,4] the string
shape fluctuations we define as $X^{\mu} = \bar{X}^{\mu} +
\eta^{\mu}(\bar{X})$, where $\eta^{\mu}(\bar{X})$ describes
fluctuations around the fixed surface $S(L)$ swept by the shape $L$
and $\bar{X}^{\mu}$ is a point upon this surface. It is assumed that
at the boundary $\partial S(L)$ of the surface $S(L)$ the fluctuating
field $\eta^{\mu}(\bar{X})$ vanishes,
i.e. $\eta^{\mu}(\bar{X})|_{\partial\,S(L)} = 0$ [3,4]. The Wilson
loop averaged over the string shape fluctuations we define as $\langle
\!\langle W({\cal C}) \rangle\!  \rangle =\langle W[X] \rangle$, where
\begin{eqnarray}\label{label4.1}
\hspace{-0.3in}\langle W[X] \rangle = \int {\cal D}\eta\,\exp\Big\{
i\int d^4x\Big[{\cal L}_{\rm eff~string}(x; \bar{X} + \eta) +
\frac{1}{2}\,\langle C_{\mu\nu}(x; \bar{X} + \eta) \rangle \,{\cal
G}^{\mu\nu}(x)\Big]\Big\}.
\end{eqnarray}
For the fluctuations around the static straight string of the length
$L$ stretched between static quark and antiquark along the $z$--axis
the integration over $\eta$--fields gives at $L\to \infty$ [3]
\begin{eqnarray}\label{label4.2}
\langle W[X] \rangle = {\displaystyle e^{\displaystyle - \sigma\,L\,T
- \delta\,V(L)\,T}},
\end{eqnarray}
where $\delta\,V(L) = -\alpha_{\rm string}/L$ and $\alpha_{\rm
string}$ is the universal coupling constant equal to $\alpha_{\rm
string} = \pi/12$ and $\alpha_{\rm string} = \pi/3$ for opened [3,4]
and closed [3] strings, respectively.

\section{Conclusion}

\hspace{0.2in} We have calculated the averaged value of the Wilson
loop in the dual representation defined for the dual Higgs model with
dual Dirac strings with Abelian $U(1)$ gauge symmetry.  We have shown
that the averaged value of the Wilson loop in the dual representation
obeys the area--law falloff testifying confinement of quarks and
antiquarks. We have found that the main contribution to the area--law
falloff comes from the quantum field fluctuations around the Abrikosov
flux line induced by a dual Dirac string in the superconducting vacuum
of the symmetry broken phase. The contribution of dual Dirac string
shape fluctuations leads to the appearance of a Coulomb--like
potential with a universal coupling constant equal to $\alpha_{\rm
string} = \pi/12$ and $\alpha_{\rm string} = \pi/3$ for opened [3,4]
and closed [3] strings, respectively.

\section*{Acknowledgement}

\hspace{0.2in} Discussions with Manfried Faber, Andrei Ivanov and Oleg
Borisenko are appreciated. One of the authors (N. I. Troitskaya) is
grateful to the staff of Atomic and Nuclear Institute of the Austrian
Universities and especially to Manfried Faber for financial support
and warm hospitality extended to her during her stay at Vienna when
this work was completed.

\end{document}